\documentclass[preprint, superscriptaddress, showpacs,preprintnumbers,amsmath,amssymb]{revtex4}
\usepackage{graphicx}

\begin{document}

\thispagestyle{empty}

\title{Thermal Casimir-Polder interaction of different
atoms with graphene}

\author{M.~Chaichian,${}^1$ G.~L.~Klimchitskaya,${}^2$
V.~M.~Mostepanenko,${}^2$ and A.~Tureanu}
\affiliation{Department of Physics, University of Helsinki,
P.O.~Box 64, FIN-00014, Helsinki, Finland\\
${}^{\it 2}$Central Astronomical Observatory
at Pulkovo of the Russian Academy of Sciences,
St.Petersburg, 196140, Russia}

\begin{abstract}
The thermal correction to the energy of Casimir-Polder interaction
of atoms with a suspended graphene membrane described by the Dirac
model is investigated. We show that a major impact on the thermal
correction is made by the size of the gap in the energy spectrum of
graphene quasiparticles. Specifically, if the
temperature is much smaller than the gap parameter (alternatively,
larger or of the order of the gap parameter), the thermal correction is
shown to be relatively small (alternatively, large).
We have calculated the free energy of the thermal Casimir-Polder
interaction of atoms of He${}^{\ast}$, Na, Rb, and Cs with
graphene described by both the hydrodynamic and Dirac models.
It is shown that in exact computations using the Dirac model,
one should use the polarization operator at nonzero temperature.
The computational results for the Casimir-Polder free energy
obtained in the framework of hydrodynamic model of graphene
are several times
larger than in the Dirac model within the separation region
below $2\,\mu$m. We conclude that the theoretical
predictions following from the two models can be reliably
discriminated in experiments on quantum reflection of different
atoms on graphene.
\end{abstract}

\maketitle

\section{Introduction}

The Casimir-Polder force acting between a microparticle and
a macrobody is caused by the existence of quantum fluctuations
of the electromagnetic field. This phenomenon is a manifestation
of the more general dispersion forces at relatively large
separations where the retardation of the electromagnetic
interaction becomes significant (at separations below a few
nanometers the electromagnetic interaction can be considered
as instantaneous and the dispersion force is usually referred to
as the van der Waals force). Quantum theories of the
van der Waals and Casimir-Polder forces were developed by
London \cite{1} and (for the case of an atom interacting with
an ideal metal plate) by Casimir and Polder \cite{2},
respectively.
For an atom interacting with a plate made of some real
material described by the frequency-dependent dielectric
permittivity, the unified theory of the van der Waals and
Casimir-Polder forces was developed by Lifshitz \cite{3}
(see also monographs \cite{4,5,6}).

Nowadays, the Casimir-Polder interaction between
different atoms and material plates (cavity walls)
has attracted
much attention in experiments on quantum reflection \cite{7,8,9}.
This is a process in which an ultracold atom is reflected by
an {\it attractive} atom-wall potential or, in other words,
it is an
above-barrier reflection. It was shown \cite{10} that quantum
reflection is very sensitive to the specific form of the
Casimir-Polder interaction. Theoretically, calculations of the
Casimir-Polder forces between various atoms and plates made of
different materials were performed \cite{11,12,13,14} on the
basis of the Lifshitz theory. The results obtained were
compared \cite{15} with those calculated using a simple
phenomenological potential \cite{8,16,17}.

Recently, special attention has been directed to the Casimir-Polder
interaction with carbon nanostructures, such as graphene, carbon
nanotubes, and fullerenes \cite{18}.
Computations were performed using the phenomenological
density-functional approach \cite{19,20,21,22}, second-order
perturbation theory \cite{23} and, for multiwalled carbon
nanotubes with at least several walls, using the Lifshitz
theory \cite{24}. For one-atom-thick carbon nanostructures,
the concept of dielectric permittivity is not immediately
applicable (it becomes applicable, for example, to thin fullerene
films and can be used to deduce the dynamic polarizabilities
of single fullerene molecules \cite{24a}).
In order to extend the Lifshitz theory to this case, the
reflection coefficients of electromagnetic oscillations on
graphene were found using the
{\it hydrodynamic model} \cite{25,26}. For this purpose,
graphene was treated as an infinitesimally thin positively
charged flat sheet, carring a homogeneous fluid with some
mass and negative charge densities.
The obtained reflection
coefficients depend on temperature only through the
Matsubara frequencies.
Calculations of the Casimir-Polder interaction in the framework
of the hydrodynamic model are presented in Refs.~\cite{27,28}
(see also \cite{6,29}).

The hydrodynamic model of graphene is only a crude approximation.
It does not take into consideration that low-energy excitations
in graphene are massless Dirac fermions except for the fact
that they move with  a Fermi velocity, rather than with the
speed of light \cite{18,30}. {}From this, it follows that
at low energies the dispersion relation for quasiparticles
is linear with respect to the momentum measured relative to
the corner of the graphene Brillouin zone.
The  {\it Dirac model} of graphene assumes that the dispersion
relation is linear at any energy. Using this assumption,
the corresponding reflection coefficients for the electromagnetic
oscillations at zero temperature have been found \cite{31}
different from those obtained with the hydrodynamic model.
The obtained coefficients depend on the polarization operator
in an external electromagnetic field calculated in the
one-loop approximation in three-dimensional spacetime.
Note that the exact value of the gap parameter $\Delta$
of quasiparticle excitations entering the polarization
operator is not known. In Ref.~\cite{32} it was shown that
Dirac model leads to much smaller values of the van der Waals
(Casimir-Polder) atom-graphene interaction than the
hydrodynamic model at separations below 100\,nm.
Keeping in mind the precision achieved in experiments on
quantum reflection of ultracold atoms \cite{7,8,9,33,34},
it was concluded that it is possible to experimentally
distinguish between the predictions of the Dirac and
hydrodynamic models of graphene.

The thermal effect in the Casimir and Casimir-Polder interactions
is a subject of debate \cite{6,29}. For the thermal
interaction between two graphene sheets it was argued
\cite{35} that the thermal regime starts at rather low
temperatures $T$ (or, respectively, short separation
distances of tens of nanometers at room temperature)
because the value of the effective temperature is determined
by the Fermi velocity $v_F$ rather than the speed of light $c$.
This conclusion was, strictly speaking, obtained in a
nonretarded regime. A fully consistent quantum-field
version of the Dirac model at nonzero temperature was
presented in Ref.~\cite{36}. In that paper the
temperature-dependent polarization operator and respective
coefficients for electromagnetic oscillations on graphene
at nonzero temperature were derived. Computations were
performed for a graphene layer with the gap parameter equal
to zero interacting with a metallic plate. For this system
the conclusion of Ref.~\cite{35} on the existence of large
thermal corrections at short separation distances was
qualitatively confirmed (with a proviso that the characteristic
distance separating zero- and high-temperature regimes depends
on the fine-structure constant rather than on the Fermi
velocity). In Ref.~\cite{37} quantum reflection of ultracold
atoms from thin films, semiconductor heterostructures and
graphene was considered at $T\neq 0$ with graphene being
described by both the hydrodynamic and Dirac models.
It was concluded that suspended graphene membranes produce
higher quantum-reflection probabilities than bulk matter.
The remark made earlier \cite{32} that already achieved
experimental precision allows discrimination between the
predictions of different models of graphene was supported.
This conclusion was obtained using the polarization operator
at zero temperature \cite{31}. Keeping in mind that
suspended graphene membranes with rather large diameter
(of $55\,\mu$m) are already available \cite{38}, it
seems pertinent to perform a full quantum-field theoretical
investigation of the thermal Casimir-Polder force between
various atoms and such membranes in the framework of
different models of graphene proposed in the literature.

In this paper, we calculate the free energy of the thermal
Casimir-Polder interaction between atoms of metastable helium
(He${}^{\ast}$), sodium (Na), rubidium (Rb), and cesium (Cs) and
a suspended graphene sheet. Graphene is described using either
the hydrodynamic model or the Dirac model and the obtained results
are compared. In the framework of the
Dirac model, the polarization operator at
nonzero temperature is employed and the dependences on the value
of the gap parameter $\Delta$ are investigated.
We demonstrate that under the condition $k_BT\ll\Delta$,
where $k_B$ is the Boltzmann constant, the thermal correction
is relatively small,
but if $\Delta\lesssim k_BT$ there are large thermal
corrections to the Casimir-Polder force. At the same time,
the magnitude of the thermal correction strongly depends on
the atom-graphene separation. The value of the Fermi velocity is
shown to be of no crucial influence on the magnitude of thermal
correction for atom-graphene system.
The novelty of this paper is that the interaction of different
atoms with graphene was investigated using the full Dirac model
at nonzero temperature with the temperature-dependent polarization
operator. In so doing it was found that the relative size of
thermal correction strongly depends on the gap parameter.
Our computations show
that experiments on quantum reflection are capable to
discriminate between the predictions of the hydrodynamic and
Dirac models of graphene. One can also conclude that in the
framework of the Dirac model it is necessary to take into
account the dependence of the polarization operator on
temperature. According to our results, the use of the
zero-temperature polarization operator in computations at
$T=300\,$K can lead to large errors in theoretical predictions
for some values of parameters.

The paper is organized as follows. In Sec.~II we present a brief
formulation of the two models of graphene, introduce the
reflection coefficients  and illustrate the limiting case of
zero temperature for the Dirac model.
Section~III contains the investigation of dependences of the
thermal correction to the Casimir-Polder energy on the gap
parameter. In Sec.~IV the distance dependences of the
Casimir-Polder interaction of different atoms used in
experiments on quantum reflection with graphene are calculated.
In Sec.~V the reader will find our conclusions and discussion.

\section{Two different models for the reflection coefficients
on graphene}

The unified expression for the van der Waals and Casimir-Polder
free energy of an  atom interacting with a planar structure is
given by the Lifshitz formula \cite{3,4,5,6}.
This formula can be expressed in terms of reflection coefficients
of the electromagnetic oscillations on this structure (in our
case on graphene) in the following way:
\begin{eqnarray}
&&
{\cal F}(a,T)=-\frac{k_BT}{8a^3}
\sum_{l=0}^{\infty}{\vphantom{\sum}}^{\prime}
\alpha(i\zeta_l\omega_c)\int_{\zeta_l}^{\infty}
\!dye^{-y}
\label{eq1} \\
&&~
\times\left\{2y^2r_{\rm TM}(i\zeta_l,y)-\zeta_l^2
\left[r_{\rm TM}(i\zeta_l,y)+r_{\rm TE}(i\zeta_l,y)
\right]\right\}.
\nonumber
\end{eqnarray}
\noindent
Here, $a$ is the separation distance between the atom and the
graphene sheet, $\alpha(i\xi_l)$ is the dynamic polarizability of
an atom calculated along the imaginary Matsubara frequencies
$\xi_l=2\pi k_BTl/\hbar$ with $l=0,\,1,\,2,\,\ldots\,$,
the dimensionless Matsubara frequencies are
$\zeta_l=\xi_l/\omega_c$, the characteristic frequency is
defined as $\omega_c=c/(2a)$, and $r_{\rm TM,TE}$ are the
reflection coefficients for two independent polarizations of
the electromagnetic field (transverse magnetic and transverse
electric, respectively).
The prime near the summation sign multiplies the term with
$l=0$ by 1/2.
Note that the dimensionless frequencies
$\zeta_l$ are functions of the separation $a$.

\subsection{Hydrodynamic and Dirac models of graphene}

As was discussed in Sec.~I, there exist two different models of
reflection coefficients for graphene. In the framework of the
hydrodynamic model the reflection coefficients take the form
\cite{25,26,27}
\begin{eqnarray}
&&
r_{\rm TM}(i\zeta_l,y)\equiv r_{\rm TM}^{(h)}(i\zeta_l,y)=
\frac{\tilde{K}y}{\tilde{K}y+\zeta_l^2},
\nonumber \\
&&
r_{\rm TE}(i\zeta_l,y)\equiv r_{\rm TE}^{(h)}(i\zeta_l,y)=
-\frac{\tilde{K}}{\tilde{K}+y}.
\label{eq2}
\end{eqnarray}
\noindent
Here, $\tilde{K}=2aK$ and $K=6.75\times 10^{5}\,\mbox{m}^{-1}$
is the characteristic wave number of the graphene sheet which
corresponds to the frequency
$\omega_{K}=cK=2.02\times 10^{14}\,$rad/s.
The dimensionless variable $y$ is connected with the projection
of the wave vector on the graphene sheet $k_{\bot}$ by the
equation $y=2a(k_{\bot}^2+\xi_l^2/c^2)^{1/2}$.
As was mentioned in Sec.~I, the hydrodynamic model does not
take into consideration that the dispersion relation for
quasiparticles in graphene is linear with respect to
momentum. It should be mentioned also that the parameter
$K$ in the hydrodynamic reflection coefficients (\ref{eq2})
is temperature-independent.

In the framework of the Dirac model the reflection
coefficients are expressed in terms of the components
of the polarization tensor \cite{36}
\begin{eqnarray}
&&
r_{\rm TM}(i\zeta_l,y)=
\frac{y\tilde{\Pi}_{00}}{y\tilde{\Pi}_{00}+
2(y^2-\zeta_l^2)},
\label{eq3} \\
&&
r_{\rm TE}(i\zeta_l,y)=
-\frac{(y^2-\zeta_l^2)\tilde{\Pi}_{tr}-
y^2\tilde{\Pi}_{00}}{(y^2-\zeta_l^2)(\tilde{\Pi}_{tr}
+2y)-y^2\tilde{\Pi}_{00}},
\nonumber
\end{eqnarray}
\noindent
where the dimensionless components $\tilde{\Pi}_{00,tr}$ are
connected with the dimensional ${\Pi}_{00,tr}$ by the equation
$\tilde{\Pi}_{00,tr}(i\zeta_l,y)=(2a/\hbar){\Pi}_{00,tr}(i\zeta_l,y)$.
The explicit expressions for the components of the polarization
operator at nonzero temperature were obtained in Ref.~\cite{36}.
In terms of our dimensionless variables the component
$\tilde{\Pi}_{00}$ can be identically represented in the
following form:
\begin{eqnarray}
&&
\tilde{\Pi}_{00}(i\zeta_l,y)=8\alpha(y^2-\zeta_l^2)
\int_{0}^{1}dx\frac{x(1-x)}{\left[{\tilde{\Delta}}^2+
x(1-x)f(\zeta_l,y)\right]^{1/2}}
+\frac{8\alpha}{{\tilde{v}}_F^2}\int_{0}^{1}dx
\label{eq4} \\
&&
~\times
\left\{\vphantom{\frac{{\tilde{\Delta}}^2+\zeta_l^2x(1-x)}{\left[{\tilde{\Delta}}^2+
x(1-x)f(\zeta_l,y)\right]^{1/2}}}
\frac{\tau}{2\pi}\ln\left[1+2\cos(2\pi lx)e^{-g(\tau,\zeta_l,y)}
+e^{-2g(\tau,\zeta_l,y)}\right]
-\frac{\zeta_l}{2}(1-2x)
\frac{\sin(2\pi lx)}{\cosh{g(\tau,\zeta_l,y)}+
\cos(2\pi lx)}
\right.
\nonumber \\
&&~
\left.
+\frac{{\tilde{\Delta}}^2+\zeta_l^2x(1-x)}{\left[{\tilde{\Delta}}^2+
x(1-x)f(\zeta_l,y)\right]^{1/2}}\,
\frac{\cos(2\pi lx)+e^{-g(\tau,\zeta_l,y)}}{\cosh{g(\tau,\zeta_l,y)}+
\cos(2\pi lx)}
\right\}.
\nonumber
\end{eqnarray}
\noindent
Here, $\alpha=e^2/(\hbar c)\approx 1/137$ is the fine-structure
constant, $\tilde{\Delta}=\Delta/(\hbar\omega_c)$ is the dimensionless gap
parameter, $\tau=4\pi ak_BT/(\hbar c)=\zeta_l/l$,
${\tilde{v}}_F=v_F/c$ is the dimensionless Fermi velocity,
the chemical potential $\mu$ is assumed to be equal to zero, and the
dimensionless functions $f$ and $g$ are defined as
\begin{eqnarray}
&&
f(\zeta_l,y)={\tilde{v}}_F^2y^2+(1-{\tilde{v}}_F^2)\zeta_l^2,
\label{eq5} \\
&&
g(\tau,\zeta_l,y)=\frac{2\pi}{\tau}
\left[{\tilde{\Delta}}^2+x(1-x)f(\zeta_l,y)\right]^{1/2}.
\nonumber
\end{eqnarray}
\noindent
In a similar way, for the sum of two spatial components of the
polarization tensor $\tilde{\Pi}_{tr}$ in our dimensionless
variables one obtains
\begin{eqnarray}
&&
\tilde{\Pi}_{tr}(i\zeta_l,y)=8\alpha[y^2+f(\zeta_l,y)]
\int_{0}^{1}dx\frac{x(1-x)}{\left[{\tilde{\Delta}}^2+
x(1-x)f(\zeta_l,y)\right]^{1/2}}
+\frac{8\alpha}{{\tilde{v}}_F^2}\int_{0}^{1}dx
\label{eq6} \\
&&
~\times
\left\{\vphantom{\frac{{\tilde{\Delta}}^2+\zeta_l^2x(1-x)}{\left[{\tilde{\Delta}}^2+
x(1-x)f(\zeta_l,y)\right]^{1/2}}}
\frac{\tau}{2\pi}\ln\left[1+2\cos(2\pi lx)e^{-g(\tau,\zeta_l,y)}
+e^{-2g(\tau,\zeta_l,y)}\right]
\right.
\nonumber \\
&&~
-\frac{\zeta_l(1-2{\tilde{v}}_F^2)}{2}(1-2x)
\frac{\sin(2\pi lx)}{\cosh{g(\tau,\zeta_l,y)}+
\cos(2\pi lx)}
\nonumber \\
&&~
\left.
+\frac{{\tilde{\Delta}}^2+x(1-x)[(1-{\tilde{v}}_F^2)^2\zeta_l^2-
{\tilde{v}}_F^4y^2]}{\left[{\tilde{\Delta}}^2+
x(1-x)f(\zeta_l,y)\right]^{1/2}}\,
\frac{\cos(2\pi lx)+e^{-g(\tau,\zeta_l,y)}}{\cosh{g(\tau,\zeta_l,y)}+
\cos(2\pi lx)}
\right\}.
\nonumber
\end{eqnarray}

It is easily seen that in the limiting case of zero temperature
($T\to 0,\,\,\tau\to 0$) we have $g\to\infty$ and the components
of the polarization operator (\ref{eq4}) and (\ref{eq6}) become
much simpler
\begin{eqnarray}
&&
\tilde{\Pi}_{00}(i\zeta,y)=\alpha
\frac{y^2-\zeta^2}{f(\zeta,y)}\tilde{\Phi}_{00}(i\zeta,y),
\label{eq7} \\
&&
\tilde{\Pi}_{tr}(i\zeta,y)=\alpha
\frac{y^2+f(\zeta,y)}{f(\zeta,y)}\tilde{\Phi}_{00}(i\zeta,y),
\nonumber
\end{eqnarray}
\noindent
where
\begin{equation}
\tilde{\Phi}_{00}(i\zeta,y)=8\sqrt{f(\zeta,y)}
\int_{0}^{1}dx\frac{x(1-x)}{\left[
\frac{{\tilde{\Delta}}^2}{f(\zeta,y)}+x(1-x)\right]^{1/2}}
\label{eq8}
\end{equation}
\noindent
and $\zeta$ is now the continuous dimensionless frequency.
Calculating the integral in Eq.~(\ref{eq8}) we arrive at
\begin{equation}
\tilde{\Phi}_{00}(i\zeta,y)=4\tilde{\Delta}+2\sqrt{f(\zeta,y)}
\left[1-4\frac{{\tilde{\Delta}}^2}{f(\zeta,y)}\right]\arctan
\frac{\sqrt{f(\zeta,y)}}{2\tilde{\Delta}}.
\label{eq9}
\end{equation}
\noindent
After the substitution of Eq.~(\ref{eq7}) in  Eq.~(\ref{eq3})
the reflection coefficients at zero temperature take the form
\begin{eqnarray}
&&
r_{\rm TM}(i\zeta,y)=
\frac{\alpha y\tilde{\Phi}(\zeta,y)}{\alpha y\tilde{\Phi}(\zeta,y)
+2f(\zeta,y)}\, ,
\nonumber \\
&&
r_{\rm TE}(i\zeta,y)=
-\frac{\alpha \tilde{\Phi}(\zeta,y)}{\alpha \tilde{\Phi}(\zeta,y)
+2y}\, .
\label{eq10}
\end{eqnarray}
\noindent
These equations were obtained in Ref.~\cite{31}.

\subsection{Properties of reflection coefficients in the Dirac
model}

Now we return to the consideration of general reflection
coefficients (\ref{eq3}) in the Dirac model with the
polarization operator (\ref{eq4}) and (\ref{eq6}) defined at any nonzero
temperature. As can be seen from Eq.~(\ref{eq3}), for
$y=\zeta_l$ we have
\begin{equation}
r_{\rm TM}(i\zeta_l,y)\big|_{y=\zeta_l}=1,\quad
r_{\rm TE}(i\zeta_l,y)\big|_{y=\zeta_l}=-1,
\label{eq11}
\end{equation}
\noindent
as it holds for any $\zeta_l$ and $y$ for ideal metal plane
at both zero and nonzero temperature.
At zero temperature, for a graphene sheet described by the
Dirac model, Eq.~(\ref{eq11}), however, does not hold because
in accordance with Eq.~(\ref{eq7}) the quantity
${\tilde{\Pi}}_{00}(i\zeta,y)\big|_{y=\zeta}=0$ and both
reflection coefficients in (\ref{eq3}) become indeterminate
form (i.e. zero/zero). The reflection coefficients at $T=0$ are
converted to a determinate form in  Eq.~(\ref{eq10}).
It is seen that the coefficients $r_{\rm TM,TE}$ (\ref{eq10})
at $y=\zeta$ are not equal to the limiting values of the
coefficients (\ref{eq11}) when $T\to 0$ (i.e., not equal to
1 and --1, respectively). This means that in the Dirac model
the reflection coefficients calculated under the condition
$y=\zeta_l$ are discontinuous functions of temperature at
the point $T=0$ (note that unlike the hydrodynamic model,
the reflection coefficients of the Dirac model depend on
$T$ not only through the Matsubara frequencies but also
explicitly through the polarization operator).

To illustrate the behavior of the reflection coefficients
in the Dirac model, Fig.~1(a) shows by the solid lines
$r_{\rm TM,TE}(i\zeta_1,y)$
as functions of $y\geq\zeta_1=0.49$ at $T=300\,$K,
$a=300\,$nm and $\Delta=0$ ($r_{\rm TM}$ is positive and $r_{\rm TE}$ is
negative).
As can be seen in Fig.~1(a), $r_{\rm TM}$ decreases with
decreasing $y$ and abruptly jumps to unity in the vicinity
of $y=\zeta_1$. In Fig.~1(b) the same is shown in an
enlarged scale. The dashed line in Fig.~1(b) shows the
coefficient $r_{\rm TM}(i\zeta_1,y)$ calculated with the
polarization operator at zero temperature [i.e., by
substituting ${\tilde{\Pi}}_{00}(i\zeta_1,y)$ from
Eq.~(\ref{eq7}) in Eq.~(\ref{eq3}) instead of using the
operator at $T=300\,$K defined in Eq.~(\ref{eq4})].
{} From Fig.~1(b) it is seen that at $l=1$ the use of the
polarization operator ta $T=0$ leads to almost the same
values of the TM reflection coefficient as the use of
the operator at $T=300\,$K (the relative difference
between the solid and dashed lines for almost all values
of $y$ is of about 4\%). Significant difference between
the two calculation methods arises only within a very narrow
interval from $\zeta_1$ to $\zeta_1+10^{-7}$.
For $l\geq 2$ all differences under discussion become even
smaller.

The solid line illustrating the behavior of the reflection
coefficient $r_{\rm TE}(i\zeta_1,y)$  as a function of $y$
in Fig.~1(a) almost coincides
with the horizontal coordinate axis in the scales used.
In the close proximity of $y=\zeta_1$, the reflection coefficient
abruptly jumps to minus unity. On an enlarged scale the
behavior of $r_{\rm TE}(i\zeta_1,y)$ as a function of $y$ is
shown in Fig.~1(c). {}From the comparison of Fig.~1(b) and
Fig.~1(c), one can conclude that
$|r_{\rm TE}(i\zeta_1,y|\ll|r_{\rm TM}(i\zeta_1,y|$
for all $y$ with exception of only a very narrow vicinity of
the point $y=\zeta_1$. Note also that the values of
$r_{\rm TE}(i\zeta_1,y)$ calculated using the polarization
operators at zero temperature and at $T=300\,$K are
indistinguishable in the scales of both Fig.~1(a) and 1(c).
The same holds for $l\geq 2$. Thus, the use of the polarization
operator (\ref{eq7}) instead of (\ref{eq4}) and (\ref{eq6})
in calculations of $r_{\rm TE}(i\zeta_1,y)$ with $l\geq 1$
leads to even smaller errors than for $r_{\rm TM}(i\zeta_1,y)$.

In Ref.~\cite{36} it was proposed to use the reflection coefficients
(\ref{eq10}) taken at zero temperature in all terms of the Lifshitz
formula with $l\geq1$ and restrict the application of the exact
reflection coefficients (\ref{eq3}), (\ref{eq4}) and (\ref{eq6})
to only the zero-frequency term $l=0$. In our computations
performed below we determine the accuracy of this prescription.

\section{Dependence of the thermal correction on a gap
parameter}

In this section we calculate the free energy of thermal
Casimir-Polder atom-graphene interaction and the thermal
correction to the Casimir-Polder energy in the framework
of the Dirac model of graphene with different values
of the gap parameter $\Delta$.
The exact value of $\Delta$ is yet unknown.
The upper bound on $\Delta$ is approximately equal to 0.1\,eV,
but it might be also much smaller \cite{18}.
 As an atomic system
interacting with graphene, we choose an atom of
metastable helium He${}^{\ast}$ often used in
experiments on quantum reflection \cite{17}.
To perform computations using the Lifshitz formula
(\ref{eq1}), one needs the dynamic atomic polarizability
of He${}^{\ast}$ as a function of the imaginary frequency.
For many atoms the dynamic polarizability can be expressed
with sufficient precision, using the single-oscillator
model
\begin{equation}
\alpha(i\omega_c\zeta_l)=
\frac{\alpha(0)}{1+(\omega_c^2/\omega_0^2)\zeta_l^2},
\label{eq12}
\end{equation}
\noindent
where $\alpha(0)$ is the static polarizability and
$\omega_0$ is the characteristic absorption frequency.
Specifically, for He${}^{\ast}$ we have
$\alpha(0)=\alpha^{{\rm He}^{\ast}}(0)=315.63\,\mbox{a.u.}
=4.678\times 10^{-29}\,\mbox{m}^3$ (where one atomic unit
of polarizability is equal to
$1.482\times 10^{-31}\,\mbox{m}^3$) and
$\omega_0=\omega_0^{{\rm He}^{\ast}}=1.18\,\mbox{eV}=
1.794\times 10^{15}\,$rad/s \cite{39}.
Note that the use of highly accurate dynamic atomic
polarizabilities (see, for instance, the polarizability of
He${}^{\ast}$ determined with a relative error $10^{-6}$
\cite{40a}) lead to only small relative deviations from the
results obtained using Eq.~(\ref{eq12}). For example, for
He${}^{\ast}$ atom near an Au wall these deviations decrease
from 3.9\% at $a=10\,$nm to 0.03\% at $a=1\,\mu$m \cite{11,13}.
Our computations show that for graphene the contribution of the
term with $l=0$ in Eq.~(\ref{eq1}) is dominant even at short
separations, i.e., the contribution of the static atomic
polarizability is of most importance. Because of this, for
graphene the single-oscillator model leads to even more
accurate results than for metallic walls.

\subsection{Casimir-Polder free energy as a function of
temperature}

We have substituted Eqs.~(\ref{eq3})--(\ref{eq6}) and
(\ref{eq12}) in Eq.~(\ref{eq1}) and performed computations
of the Casimir-Polder free energy ${\cal F}$ as a function
of temperature at atom-graphene separation $a=1\,\mu$m
for the values of a gap parameter equal to $\Delta=0.1$,
0.05, 0.01\,eV and for $\Delta=0$.
The computational results for the quantity $a^4|{\cal F}|$ are
presented in Fig.~2, where the four lines from the lowest
to highest correspond to the decreasing values of $\Delta$
(the lowest line is for $\Delta=0.1\,$eV).
As can be seen in Fig.~2, the characteristic behavior of
the free energy differs significantly for different gap
parameters. At $T=0$ the values of the Casimir-Polder energy
$E(a)={\cal F}(a,0)$ depend heavily on $\Delta$, whereas
at $T=300\,$K there is only a slight dependence of the
Casimir-Polder free energy ${\cal F}(a,T)$ on $\Delta$.
This allows estimation of $\Delta$ from the comparison
between the measurement data and computational results in the
region of moderate temperatures from 100 to 150\,K.
Furthermore, the larger is $\Delta$, the wider is the
temperature region where ${\cal F}$ remains constant
with the increase of temperature. In such temperature
regions the thermal correction to the Casimir-Polder
energy is negligibly small. Below we discuss this point
in more detail.

To check quantitatively an accuracy of the prescription
\cite{36} that in all terms of the Lifshitz formula with
$l\geq 1$ one can use the polarization operator at $T=0$,
we repeated the same computation as above, but this time with
the operators (\ref{eq4}) and (\ref{eq6}) for $l=0$ and
(\ref{eq7}) for all $l\geq 1$. The obtained results cannot
be distinguished visually from those shown in Fig.~2.
The largest deviations between the computational results
obtained using different calculation procedures (with the
polarization operators found at $T\neq 0$ for all $l$
and for only $l=0$) hold with the gap parameter
$\Delta=0$. In this case, the magnitudes of the free energy
obtained using the prescription are smaller than those in
full computations by 0.6\%, 0.3\%, 0.1\%, and 0.06\%
at $T=300\,$K and separations $a=50$, 200, 500, and
1000\,nm, respectively. At $a=3\,\mu$m the relative
difference between the computational results obtained
using the two procedures is as small as 0.002\%.
One can conclude that the prescription of Ref.~\cite{36}
leads to very accurate results and can be used in
subsequent computations.

\subsection{Thermal correction as a function of separation}

Now we calculate the thermal correction to the Casimir-Polder
energy in the interaction of He${}^{\ast}$ atom with graphene
as a function of separation. The relative thermal correction
at a temperature $T$ is defined as
\begin{equation}
\delta_T{\cal F}(a,T)=
\frac{{\cal F}(a,T)-{\cal F}(a,0)}{{\cal F}(a,0)}.
\label{eq13}
\end{equation}
\noindent
The computations are performed at $T=300\,$K using Eq.~(\ref{eq1})
with full polarization operator for $l=0$ and zero-temperature
operators for $l\geq 1$. The computational results within the
separation region from 10 to 500\,nm are shown in Fig.~3(a) by
the six solid lines from the highest to the lowest corresponding
to the values of the gap parameter $\Delta=0.1$, 0.05, 0.025,
0.01, 0.001, and 0\,eV, respectively. Note that the two lowest
lines obtained for two smallest values of the gap parameter
are almost coinciding. It is interesting that at $T=300\,$K
the magnitude of the relative thermal correction at each
separation increases monotonically with increasing $\Delta$.
The largest thermal correction is achieved for $\Delta=0.1\,$eV.
At $T=300\,$K the same properties hold at shorter separation
distances below 100\,nm. This separation region is ahown in an
enlarged scale in Fig.~3(b).

As suggested by Fig.~2, the monotonous dependence of the relative
thermal correction on $\Delta$ is not universal and does not hold
at any temperature. As an example we have computed the thermal
correction to the energy of He${}^{\ast}$-graphene interaction at
$T=100\,$K. This is the maximum temperature until which the free
energy is nearly constant for $\Delta=0.1\,$eV.
The computational results within the separation region from 10 to
500\,nm are shown in Fig.~4(a) by the three solid lines from the
lowest to the highest corresponding to the following values of
the gap parameter: $\Delta=0.1$, 0.05, and 0.025\,eV.
As is seen in Fig.~4(a), for $\Delta=0.1\,$eV the thermal
correction is very small over the entire separation region.
This is in accordance with the computational results shown in
Fig.~2. The computational results for the same thermal correction
at $T=100\,$K, but with the values of the gap parameter equal to
$\Delta=0.025$, 0.01, 0.001, and 0\,eV, are shown in Fig.~4(b)
by the solid lines
from the highest to the lowest line, respectively.
For illustration purposes, the line with $\Delta=0.025\,$eV
is reproduced in both Fig.~4(a) and 4(b).
{}From the comparison of Fig.~4(a) with Fig.~4(b) one can see
that at $T=100\,$K the dependence of the thermal correction
on $\Delta$ is not monotonous. {}From Fig.~4(b) it can be seen also
that there is a noticeable difference between the thermal
corrections for a graphene with $\Delta=0.001\,$eV and with
$\Delta=0$. Similar results are obtained for other atoms.

One can conclude that the size of thermal correction to the
Casimir-Polder interaction of an atom with a graphene sheet
depends essentially on the size of the gap in the spectrum of
graphene quasiparticles. {}From our computations it follows
that if the inequality $k_BT\ll\Delta$ is satisfied with a
large safety margin, then the relative thermal correction is
small (note that $T=100\,$K corresponds to $k_BT=0.0083\,$eV
to be compared with $\Delta=0.1\,$eV). On the contrary,
if $\Delta\lesssim k_BT$, then the thermal correction is large
(in so doing, the thermal correction can be also large for $k_BT$
smaller but not much smaller than $\Delta$).
Keeping in mind that the value of $\Delta$ for graphene is
not yet known \cite{18}, the predictions of the Dirac model concerning
the size of thermal correction in atom-graphene interaction
remain uncertain.

\section{Distance dependence for the Casimir-Polder interaction
of different atoms with graphene}

We come now to the Casimir-Polder interaction of graphene
with atoms of He${}^{\ast}$, Na, Rb, and Cs at room temperature
$T=300\,$K but at different separation distances, as is of
interest for experiments on quantum reflection. Here, we compare
the computational results obtained using both the hydrodynamic and
Dirac models for graphene.

\subsection{Atom of metastable helium}

 We start with an atom of He${}^{\ast}$
and compute the Casimir-Polder free energy (\ref{eq1}) with the
reflection coefficients (\ref{eq2}) of the hydrodynamic model.
The computational results for $a^4|{\cal F}|$ as a function of
separation from 50\,nm to $5\,\mu$m are presented in Fig.~5(a)
by the dashed line. Then we repeated computations using
Eq.~(\ref{eq1}), but with the Dirac reflection coefficients
(\ref{eq3}), which contain the polarization operator (\ref{eq4}),
(\ref{eq6}) obtained at $T=300\,$K [as was noted in Sec.~II,
it is sufficient to use this operator only with
$l=0$ term and replace it by a more simple operator (\ref{eq7})
in all terms with $l\geq 1$]. In this case the computational
results for the gap parameter in the region from 0 to 0.01\,eV
are shown by the solid line. Note that even the use of larger
gap parameters up to 0.1\,eV leads to almost imperceptible shift
of the solid line in Fig.~5(a). This is explained by the fact
that at $T=300\,$K there are only minor differences between the
Casimir-Polder free energies computed with different $\Delta$
(see Fig.~2).

In Fig.~5(a)we also plot by the dotted line the computational
results obtained using the Lifshitz formula (\ref{eq1}) and
the Dirac model with $\Delta=0$ at zero temperature [i.e.,
using the polarization operator (\ref{eq7}) with all Matsubara
terms with $l\geq 0$]. As can be seen from Fig.~5(a), the
computational results shown by the solid line are by factors of
1.27, 1.29, and 1.29 larger than those shown by the dotted line
at separations 1, 3, and $5\,\mu$m, respectively.
This confirms that in order to perform precise computations at
$T=300\,$K, one should use the polarization operator at the same
temperature (at least in the zero-frequency contribution of
the Lifshitz formula). {}From Fig.~5(a) it is also seen that the
predictions of the hydrodynamic model at the separations 0.2, 0.5,
1.0, 1.5, 2, and $3\,\mu$m (the dashed line) are by factors of
3.83, 2.96, 2.07, 1.62, 1.36, and 1.12 larger than the predictions
of the Dirac model (the solid line).
This allows reliable discrimination between theoretical
predictions of the two models by the measurement data
of experiments on quantum reflection. (Note that in Ref.~\cite{32},
 where computations were performed at $a<100\,$nm using the
 polarization operator at $T=0$, the predictions of the Dirac
 model were overestimated by approximately a factor of 1.5 due to
 an error in the computer program. As a result, the differences
 between the predictions of two models at short separations were
underestimated.)

{}From the comparison of solid and dotted lines in Fig.~5(a), we
have already found errors arising from the use of polarization
operator at $T=0$ in all terms of the Lifshitz formula
(an underestimation of $|{\cal F}|$ by approximately a factor of 1.3).
This conclusion was obtained, however, from the zero-temperature
polarization operator with $\Delta=0$ (as was demonstrated above,
computational results with the $T$-dependent polarization operator
are not sensitive to the value of $\Delta$ at $T=300\,$K).
To investigate the role of the gap parameter in computations using
the polarization operator at $T=0$, in Fig.~5(b) we plot
$a^4|{\cal F}|$ as a function of separation computed with the
full operator (\ref{eq4}), (\ref{eq6}) (the solid line) and with
the operator (\ref{eq7}) for $\Delta=0$, 0.001, and 0.01\,eV
(the dotted lines from the highest to the lowest, respectively).
Note that the solid and the highest dotted lines reproduce the
respective lines in Fig.~5(a) at separations from 50\,nm to
$1\,\mu$m. As can be seen from Fig.~5(b), for nonzero $\Delta$
the underestimation of $|{\cal F}|$ when using the polarization
operator at $T=0$ is much larger than for $\Delta=0$. Thus,
at $a=1\,\mu$m this underestimation is by the factors of 1.83 and
6.2 for the gap parameter $\Delta=0.001$ and 0.01\,eV,
respectively. This adds importance to the use of full
temperature-dependent polarization  operator (\ref{eq4}), (\ref{eq6})
in computations performed for subsequent comparison with the
experimental data.

\subsection{Atoms of sodium, rubidium and cesium}

The Casimir-Polder free energy for other atoms used in
experiments on quantum reflection can be computed in a similar
way. For Na the dynamic polarizability can be presented by
Eq.~(\ref{eq12}) with
$\alpha^{\rm Na}(0)=162.68\,\mbox{a.u.}=
2.411\times 10^{-29}\,\mbox{m}^3$ and
$\omega_0^{\rm Na}=2.14\,\mbox{eV}=3.25\times 10^{15}\,$rad/s
\cite{40}.
The computational results for the quantity $a^4|{\cal F}|$ for
Na-graphene interaction at $T=300\,$K are presented in Fig.~6(a)
as function of separation by the dashed line (the hydrodynamic
model), the solid line [the Dirac model with the $T$-dependent polarization
operator (\ref{eq4}), (\ref{eq6})], and by the dotted line
 [the Dirac model with the polarization
operator (\ref{eq7}) at $T=0$, $\Delta=0$].
It can be seen that qualitatively the computational results in Fig.~6(a)
are similar to those in Fig.~5(a) for a He${}^{\ast}$ atom.
In the case of Na, however, the values of $a^4|{\cal F}|$ are
smaller than for He${}^{\ast}$ at all respective separations.
The difference between the predictions of the hydrodynamic and
Dirac models can be easily discriminated by the measurement data
of experiments on quantum reflection. Thus, at separations
0.2, 0.5, 1.0, 1.5, and $2\,\mu$m the predictions of the
hydrodynamic model are larger than the predictions of the Dirac
model with the $T$-dependent polarization operator by
factors of 4.04, 3.05, 2.10, 1.63, and 1.37, respectively.
The use of the polarization operator at $T=0$ again
underestimates the predictions of the Dirac model
(by the factors of 1.27, 1.29, and 1.29 at separations $a=1$,
3, and $5\,\mu$m, respectively).

In Figs.~6(b) and 6(c) similar results are presented for
atoms Rb and Cs interacting with graphene.
To perform computations of the Casimir-Polder free energy
using different models of graphene, we have used the
single-oscillator model (\ref{eq12}) with the following
parameters:
$\alpha^{\rm Rb}(0)=319.9\,\mbox{a.u.}=
4.73\times 10^{-29}\,\mbox{m}^3$,
$\omega_0^{\rm Rb}=5.46\,\mbox{eV}=8.3\times 10^{15}\,$rad/s
\cite{41} and
$\alpha^{\rm Cs}(0)=403.6\,\mbox{a.u.}=
5.981\times 10^{-29}\,\mbox{m}^3$ and
$\omega_0^{\rm Cs}=1.55\,\mbox{eV}=2.36\times 10^{15}\,$rad/s
\cite{40,42}.
As can be seen from the comparison of Fig.~6(b) with Fig.~6(a) and
Fig.~6(c) with Fig.~6(b), the magnitudes of the free energy for Rb
are larger than for Na and the respective magnitudes
for Cs are larger than for Rb at all
separation distances. This is explained by the fact that the
static polarizability of Cs is the largest one.
{}From Figs.~6(b) and 6(c) one can observe the same pattern
between the dashed, solid and dotted lines, obtained for
the hydrodynamic model and for the Dirac model with
$T$-dependent and $T$-independent polarization operators,
respectively, as was discussed on the basis of Figs.~5(a) and 6(a).

In Figs.~5(a) and 6(a-c), the distinction between the predictions of
the hydrodynamic model (the dashed lines) and the Dirac model with
$T$-dependent  polarization operator (the solid lines) deserves
special attention. The point is that the dashed lines demonstrate
the same qualitative behavior \cite{15} as the Casimir-Polder
free energy of atoms interacting with walls made of ordinary
real materials, such as Au or Si. By contrast, the Dirac model
for graphene predicts quite a different behavior of the free
energy, where it is nearly inverse proportional to the third
power of separation. Keeping in mind that at separations
below $2\,\mu$m the predictions of both models may differ
by a factor of 2 and even up to a factor of 4, it seems
appropriate to raise a question which model is in better
agreement with the experimental data. This question can be
answered through experiments on quantum reflection of
different atoms on suspended graphene membranes.

\section{Conclusions and discussion}

In this work we have investigated the Casimir-Polder
interaction of different atoms with graphene in the thermal
regime. Special attention was paid to the Dirac model of
graphene and to the conditions when thermal effects become
essential. We have confirmed the conclusion \cite{36} that
sufficient precision can be achieved by using the
temperature-dependent polarization operator in the
zero-frequency term of the Lifshitz formula alone and
calculating all the other terms using the polarization
operator at $T=0$. We have found that a major impact
on the thermal correction (i.e. whether
it is small or large) is
made by the size of the gap $\Delta$ in the spectrum of
graphene nanoparticles. According to our results, if the
condition $k_BT\ll\Delta$ is satisfied with a large safety
margin, the thermal correction to the Casimir-Polder
interaction of an atom with graphene is relatively small.
In future it is planed to obtain the asymptotic behavior of
the free energy under this condition analytically and to
verify the fulfilment of the Nernst heat theorem in the
Dirac model.
If, however, $\Delta\lesssim k_BT$, the thermal
correction is relatively large.

Furthermore, we have calculated the free energy of the
Casimir-Polder interaction of different atoms used in
experiments on quantum reflection (such as He${}^{\ast}$,
Na, Rb, and Cs) with a suspended graphene membrane.
All computations were performed for room temperature within
a wide separation region in the framework of two models
for graphene proposed in the literature: the hydrodynamic
and Dirac ones. It is important to note that both these
models are approximations and {\it {a} priori} it
is not possible to decide which of them provides a better
theoretical description of the Casimir-Polder force
(for example, the Drude model of metals describes correctly
the electric properties of metals and their dielectric
response to external electromagnetic fields, but is in
disagreement with the experimental data on measurements of
the Casimir force between metallic test bodies \cite{6,29}).

Our computational results allow to conclude that the
hydrodynamic and Dirac models of graphene lead to both
qualitatively and quantitatively different results for the
free energy of atom-graphene Casimir-Polder interaction.
{}On the qualitative side, we have arrived at quite different
dependences of the free energy on separation distance.
Quantitatively, the magnitudes of the free energies computed
using the two models differ by a factor of 2 and even by a
factor of 4 at different separations below $2\,\mu$m.
A difference in theoretical predictions being as large as
by this amount
assures a reliable discrimination between the hydrodynamic and
Dirac models of graphene by comparing with the experimental data
 on quantum reflection. Keeping in mind that
suspended graphene membranes have been already produced
\cite{38}, one could expect the resolution of this problem
in the immediate future.

\section*{Acknowledgments}
The financial support of the Academy of Finland
under the Projects No.\ 136539 and 140886 is gratefully
acknowledged.
 G.L.K.\ and V.M.M.\ were also partially
supported by the DFG grant BO\ 1112/21--1.


\begin{figure*}[h]
\vspace*{1.0cm}
\centerline{\hspace*{-1cm}
\includegraphics{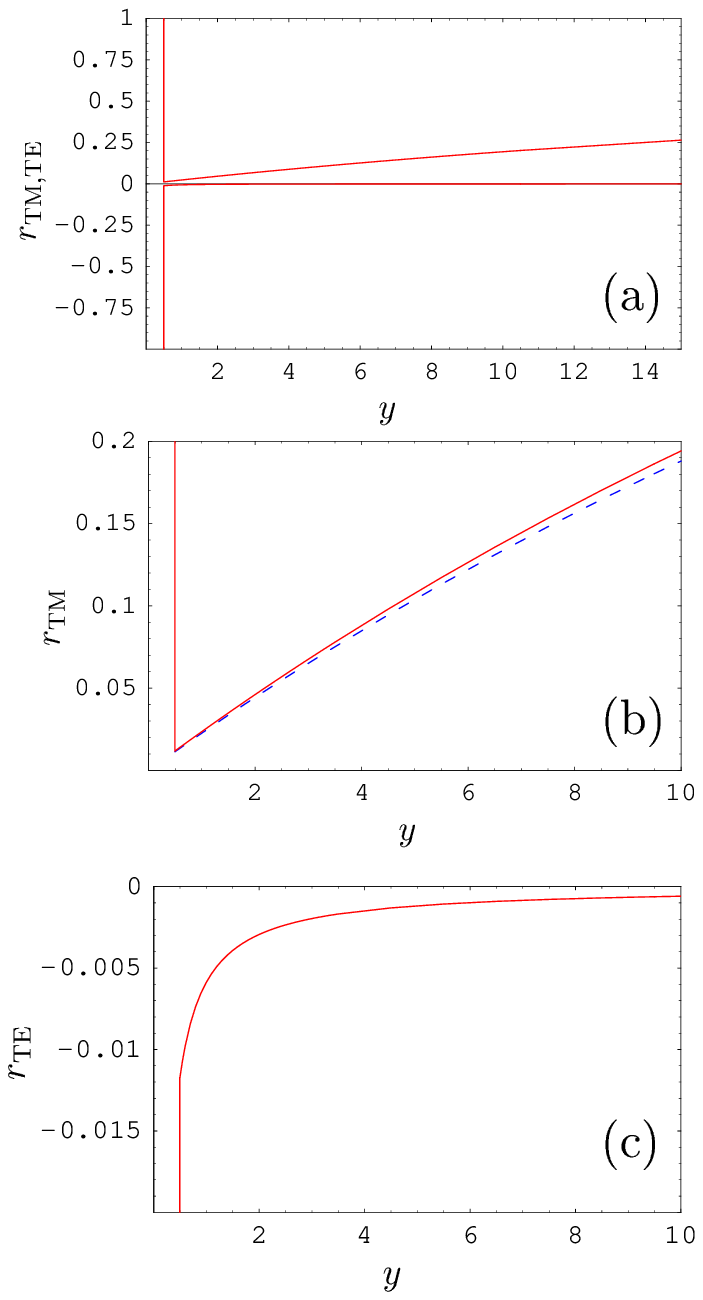}
}
\vspace*{-16.cm}
\caption{(Color online)
(a) The reflection coefficients $r_{\rm TM,TE}$ on graphene
calculated at $\zeta=\zeta_1$, $T=300\,$K and $a=300\,$nm
as functions of the dimensionless variable $y$ using the polarization
tensor at $T=300\,$K are shown by the positive- and
negative-valued solid lines, respectively.
(b) The solid and dashed lines show $r_{\rm TM}$ computed
using the polarization
tensor at $T=300\,$K  and $T=0$, respectively.
(c) The solid line shows $r_{\rm TE}$.
The scales in (b) and (c) are enlarged as compared to (a).
}
\end{figure*}
\begin{figure*}[h]
\vspace*{-13.0cm}
\centerline{\hspace*{2.0cm}
\includegraphics{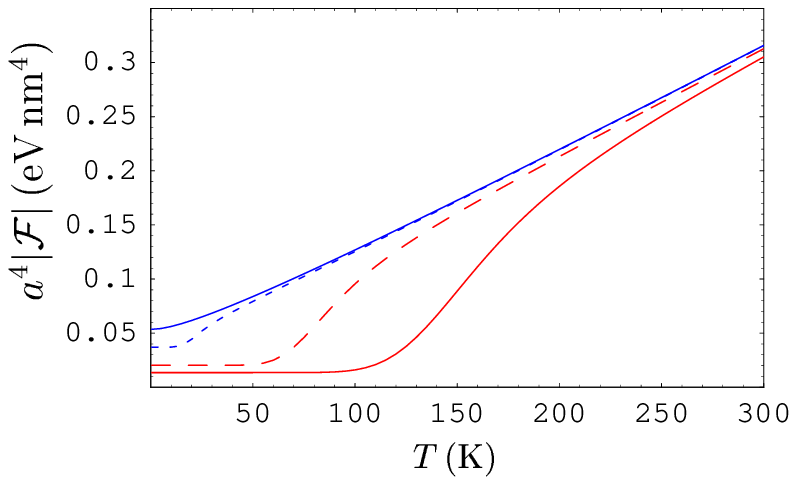}
}
\vspace*{-10.cm}
\caption{(Color online)
The Casimir-Polder free energy of He${}^{\ast}$-graphene interaction
at $a=1\,\mu$m multiplied by the fourth power of separation
is shown as a function of temperature
by the four lines from the lowest to the
highest for the gap parameter $\Delta=0.1$, 0.05, 0.01,
and 0\,eV, respectively.
}
\end{figure*}
\begin{figure*}[h]
\vspace*{-1.0cm}
\centerline{\hspace*{-1cm}
\includegraphics{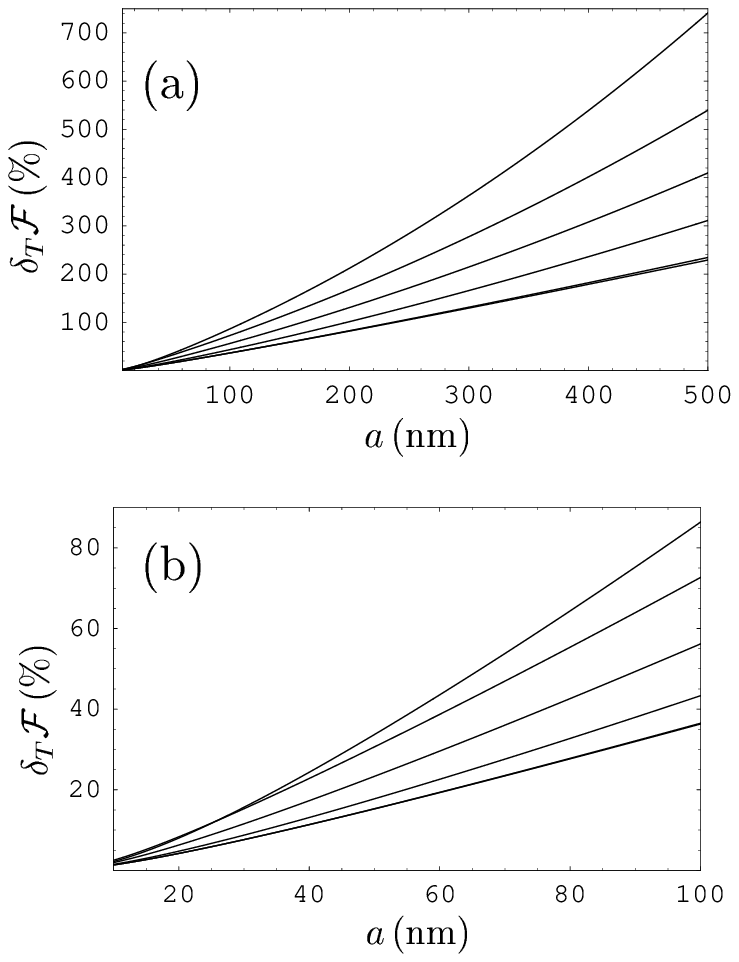}
}
\vspace*{-16.cm}
\caption{
(a) The relative thermal correction to the Casimir-Polder energy of
He${}^{\ast}$-graphene interaction
at $T=300\,$K multiplied by the fourth power of separation
is shown as a function of separation by the solid lines
from the highest to the lowest for
 the gap parameter $\Delta=0.1$, 0.05, 0.025, 0.01, 0.001,
and 0\,eV, respectively.
(b) The same is shown at separations from 10 to 100\,nm.
}
\end{figure*}
\begin{figure*}[h]
\vspace*{-1.0cm}
\centerline{\hspace*{-1cm}
\includegraphics{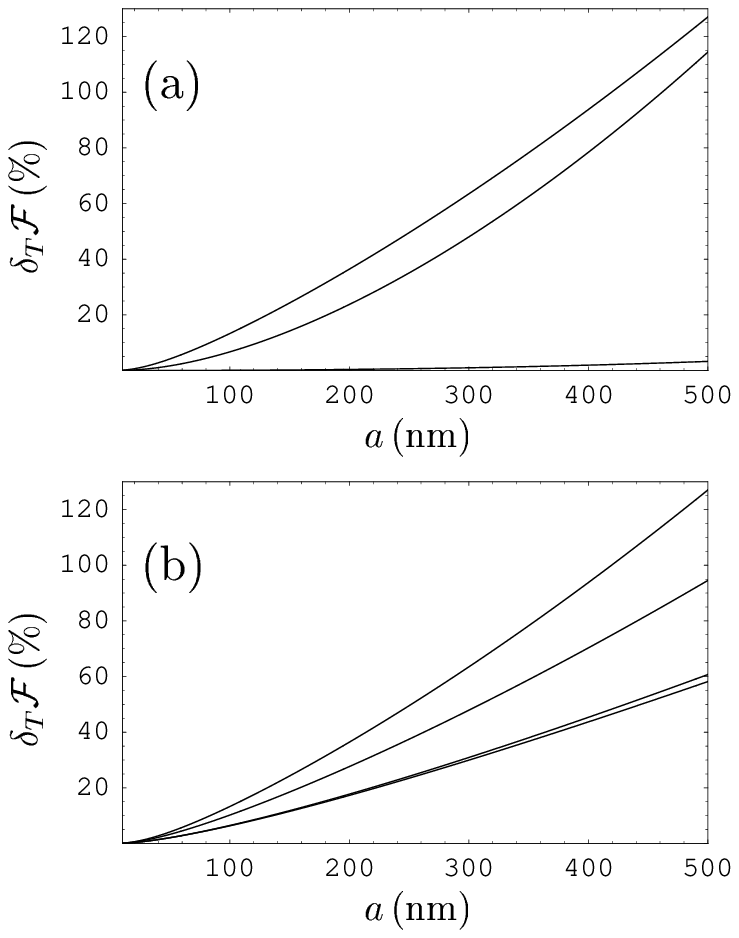}
}
\vspace*{-16.cm}
\caption{
The relative thermal correction to the Casimir-Polder energy of
He${}^{\ast}$-graphene interaction
at $T=100\,$K multiplied by the fourth power of separation
is shown as a function of separation by the solid lines (a) from
the lowest to the highest for the gap parameter $\Delta=0.1$,
0.05, and 0.025\,eV and
(b) from the highest to the lowest for
 the gap parameter $\Delta=0.025$, 0.01, 0.001,
and 0\,eV.
}
\end{figure*}
\begin{figure*}[h]
\vspace*{-1.0cm}
\centerline{\hspace*{-1cm}
\includegraphics{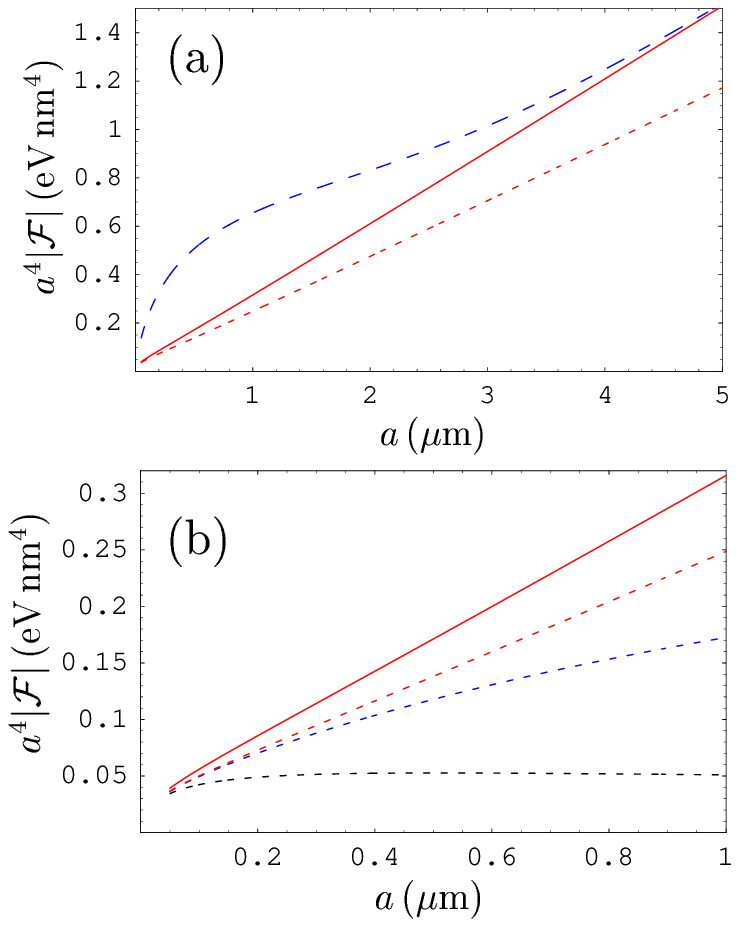}
}
\vspace*{-16.cm}
\caption{(Color online)
The Casimir-Polder free energy of He${}^{\ast}$-graphene interaction
at $T=300\,$K multiplied by the fourth power of separation
is shown as a function of separation (a) by the dashed, solid and
dotted lines using the hydrodynamic model, the Dirac model
with $T$-dependent polarization operator, and with the
polarization operator at zero temperature, respectively.
(b) The same quantity is shown where the dotted lines from the
highest to the lowest are computed using the polarization operator
at zero temperature with the gap parameter $\Delta=0$, 0.001,
and 0.01\,eV, respectively.
}
\end{figure*}
\begin{figure*}[h]
\vspace*{1.0cm}
\centerline{\hspace*{-1cm}
\includegraphics{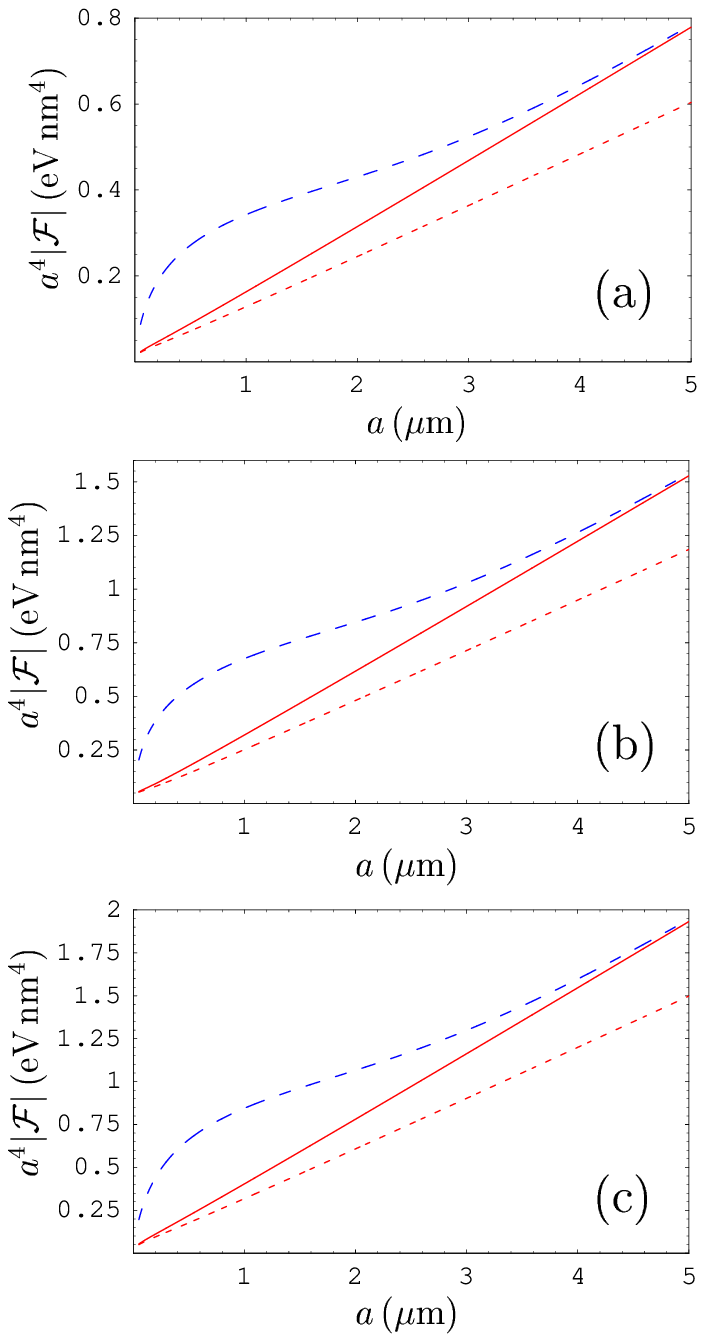}
}
\vspace*{-15.5cm}
\caption{(Color online)
The Casimir-Polder free energy of atom-graphene interaction
at $T=300\,$K multiplied by the fourth power of separation
is shown as a function of separation by the dashed, solid and
dotted lines using the hydrodynamic model, the Dirac model
with $T$-dependent polarization operator, and with the
polarization operator at zero temperature, respectively, for
atoms of (a) Na, (b) Rb, and (c) Cs.
}
\end{figure*}

\end{document}